\documentclass[prl,amsmath,twocolumn,showpacs]{revtex4}
\usepackage{amsmath,graphicx}
\usepackage{epsfig}
\begin{document}
\def\la{{\langle}}
\def\ra{{\rangle}}
\def\a{{\alpha}}

\title{Accurate "superluminal'' transmission via entanglement, superoscillations and quasi-Dirac distributions}
%
\author {D. Sokolovski$^{1,2,3} $and R. Sala Mayato$^{4}$}
\address{$^1$ Department of Chemical Physics,  University of the Basque Country, Leioa, Spain\\
$^2$ IKERBASQUE, Basque Foundation for Science\\
$^3$ School of Maths and Physics, Queen's University of Belfast, Belfast, BT7 1NN, UK\\
$^4$ F\'\i sica Fundamental II and IUdEA, Universidad de La Laguna, La Laguna, 38204 (S/C de Tenerife), Spain} 

   \begin{abstract}
   We analyse a system in which, due to entanglement between the
   spin and spatial degrees of freedom, the reduced transmitted state has
   the shape of the freely propagating pulse translated in the complex 
   co-ordinate plane. In the case an apparently ``superluminal'' advancement of the pulse the delay 
   amplitude distribution
   is found to be a peculiar approximation to the Dirac $\delta$-function, and the transmission
   coefficient exhibits a well-defined  super-oscillatory window.
   Analogies with potential tunnelling and the Wheeler's delayed choice experiment are highlighted.
\end{abstract}

%
%
\pacs{PACS number(s): 03.65.Ta, 73.40.Gk}
\maketitle
{\it Introduction.}
Quantum speed up effect known since early thirties of last century allows the maximum of a wavepacket,
transmitted across a classically forbidden region, arrive at a detector earlier than that of a freely propagating one. It has been predicted and observed for various systems such as potential barriers, semi-transparent mirrors, refraction of light, microwaves in undersized wave guides and fast-light materials  \cite{REV}, \cite{Chen}
A transmission appears ``superluminal'' if
one uses the advancement of the transmitted peak to predict the time spent by the particle, photon or electromagnetic pulse inside a barrier, scatterer or a waveguide. It is generally understood that such a prediction is incorrect since the initial pulse, usually greatly reduced, undergoes in the process a severe reshaping.

Superoscillatory functions  which, in a limited region, oscillate with a frequency outside
the support of their Fourier spectrum were introduced in \cite{BERRY1} and extensively 
studied in \cite{KEMPF}. These functions have been recently applied for improving optical resolution beyond the diffraction limit \cite{BERRY2}.
Authors of Ref. \cite{AH1} have established a connection between superoscillatory behaviour and anomalous tunnelling times (weak values) which occur in ``superluminal'' signal transmission.
The analogy between apparent ``superluminality'' and weak measurements was further explored in Ref. \cite{SMS}, and additional work on the weak values and superoscillations  can be found in \cite{TOLL}.

The purpose of this paper is to propose and analyse a system in which a similar speed up effect 
occurs due to entanglement between the spin and spatial degrees of freedom in a magnetic field
 so that the transmitted 
state is essentially a reduced copy of the freely propagating one,
with an additional coordinate shift.
 We will show that in the case of a significant advancement the delay amplitude distribution \cite{SMS} (DAD)
 is a wildly oscillatory function with a finite support $\Omega$ which approximates a Dirac $\delta$-function with support outside $\Omega$. In the momentum space, the transmission coefficient exhibits a superoscillatory behaviour within a well defined band.  We will also demonstrate that manipulations of the particle's spin states allows to choose between the regimes where the particle spend a known amount of time in the magnetic field and where, in the spirit of wave-particle duality, such a duration cannot be defined.

{\it Quantum speed up effect via entanglement.}
We consider a particle of a unit mass $\mu=1$ 
equipped with a $2K+1$-component magnetic moment (spin).
The particle is described by a wave packet with a mean momentum
$p_0$, which  propagates without spreading in the time interval of interest.
Thus, its freely propagating wave function is given by
\begin{equation}\label{1}
\la x|\Psi(t)\ra=\exp(ip_0x-ip_0^2t/2) G(x-p_0t) |a\ra
\end{equation}
where $G$ is the envelope and $|a\ra \equiv  \sum_{m=-K}^K a_m |m\ra/ \sqrt{N(a)}$
\begin{equation}\label{1}
N(a)\equiv{\sum_{m=-K}^K  |a_m|^2}
\end{equation}
is the initial state of the spin
written here in terms of its components, $m$, along the $z$-axis.
Next we make the particle pass through a constant magnetic field in the $z$-direction
created in the region  $0<x<d$
so that  the $m$-th component of the spin wave function encounters there an additional rectangular potential of the magnitude $m \omega_L$ ($\omega_L$ is the Larmor frequency), a well if $m<0$ and a 
barrier for $m>0$. If the particle is sufficiently fast, $p_0^2/2 >> K\omega_L$,  the reflection off the field's edges can be neglected, and upon traversing the field the $m$-th component of the wave function
will be advanced or lagging behind  the freely propagating pulse by 
$ m\Delta x $, $\Delta x \equiv \omega_Ld/p_0^2.$
On exit from the magnetic field the wave function 
contains a superposition of shifted wave packet's envelopes,  with shifts ranging from
$-K\Delta x$ to $K\Delta x$,
\begin{eqnarray}\label{2}
\la x|\Psi(t)\ra= 
\exp(ip_0x-ip_0^2t/2) \sum_{m=-K}^{K} a_m \times \\
\nonumber
\exp(-im\omega_L d/p_0) G(x-p_0t-m\Delta x) |m\ra/ \sqrt {N(a)}.  
\end{eqnarray} 
For simplicity we will choose an initial spin state with  $a_m =0$ for all $m>0$,  so that none of the advanced shapes enter the sum
(\ref{2}), and post-select the particle spin in some known final state $|b\ra \equiv  \sum_{m=-K}^K b_m |m\ra
/ \sqrt {N(b)}$. 
We assume that such a post-selection can be performed as measuring the projector on $|b\ra$, $\hat{P}(b)$, e.g., 
by making the particle pass through an additional polariser.
On exit from the polariser the envelope of the particle's state $\la b|\Psi(t)\ra$ becomes ($X\equiv x-p_0t$):
\begin{eqnarray}\label{6}
\tilde{G}(X) = \int_{-\infty}^\infty G(X-x')\eta(x') dx'/ \sqrt {N(a)N(b)}\quad\quad \quad
\end{eqnarray}
where
\begin{eqnarray}\label{6a}
 \eta(x) \equiv \sum_{m=0}^K \eta_m\delta(x+m\Delta x),\quad\\
 \nonumber
  \eta_m\equiv \exp(-im\omega_Ld/p_0)a_mb^*_m .
\end{eqnarray}
We wish  the envelope of the transmitted pulse to be an accurate  (although possibly reduced) copy 
of the free envelope translated in space by a distance $\a$,
\begin{eqnarray}\label{4}
\tilde{G}(x,t) \approx G(X-\a)/ \sqrt {N(a)N(b)}.
\end{eqnarray} 
Choosing $\a >0$ gives the impression that the particle has been sped up while passing through
the set up containing the the magnetic field.

\begin{figure}[h]
\includegraphics[width=9cm, angle=0]{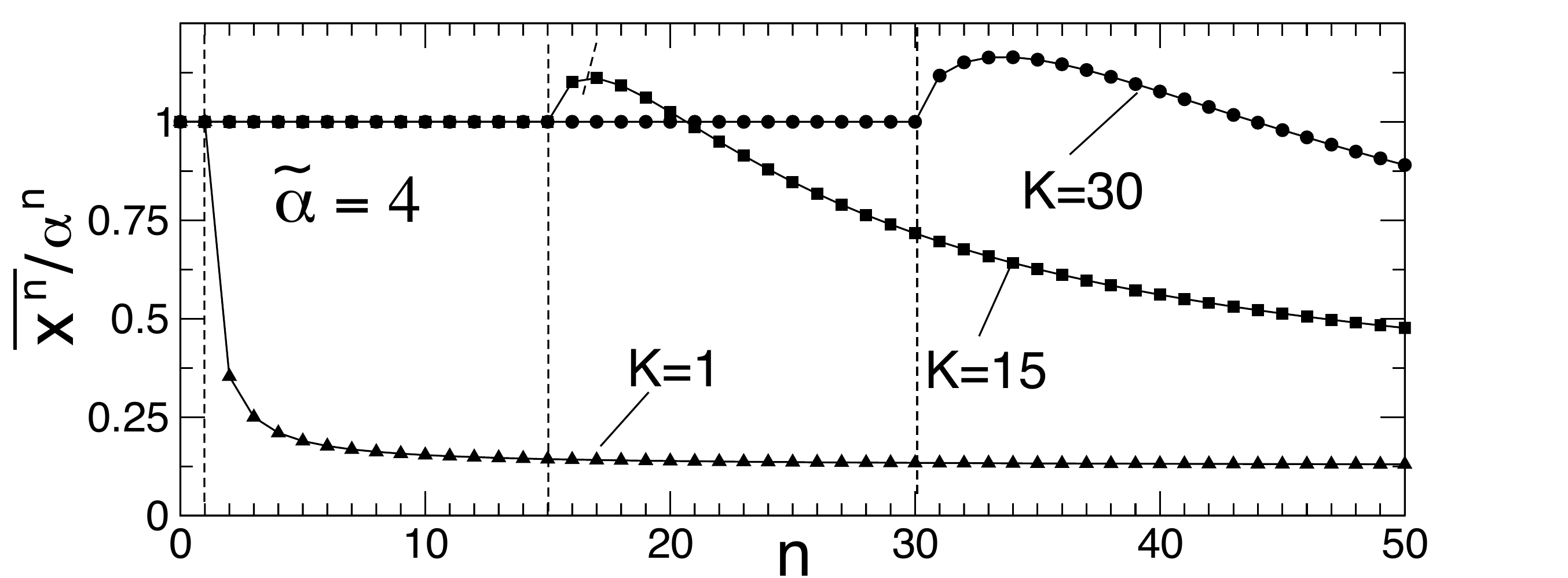}
\caption{The moments of the DAD (\ref{10}) with $K=1$, $15$ and $30$ for $\tilde{\a}\equiv \a/(K\Delta x)=4.$}
\label{FIG1}
\end{figure}

\begin{figure}[h]
\includegraphics[width=8cm, angle=0]{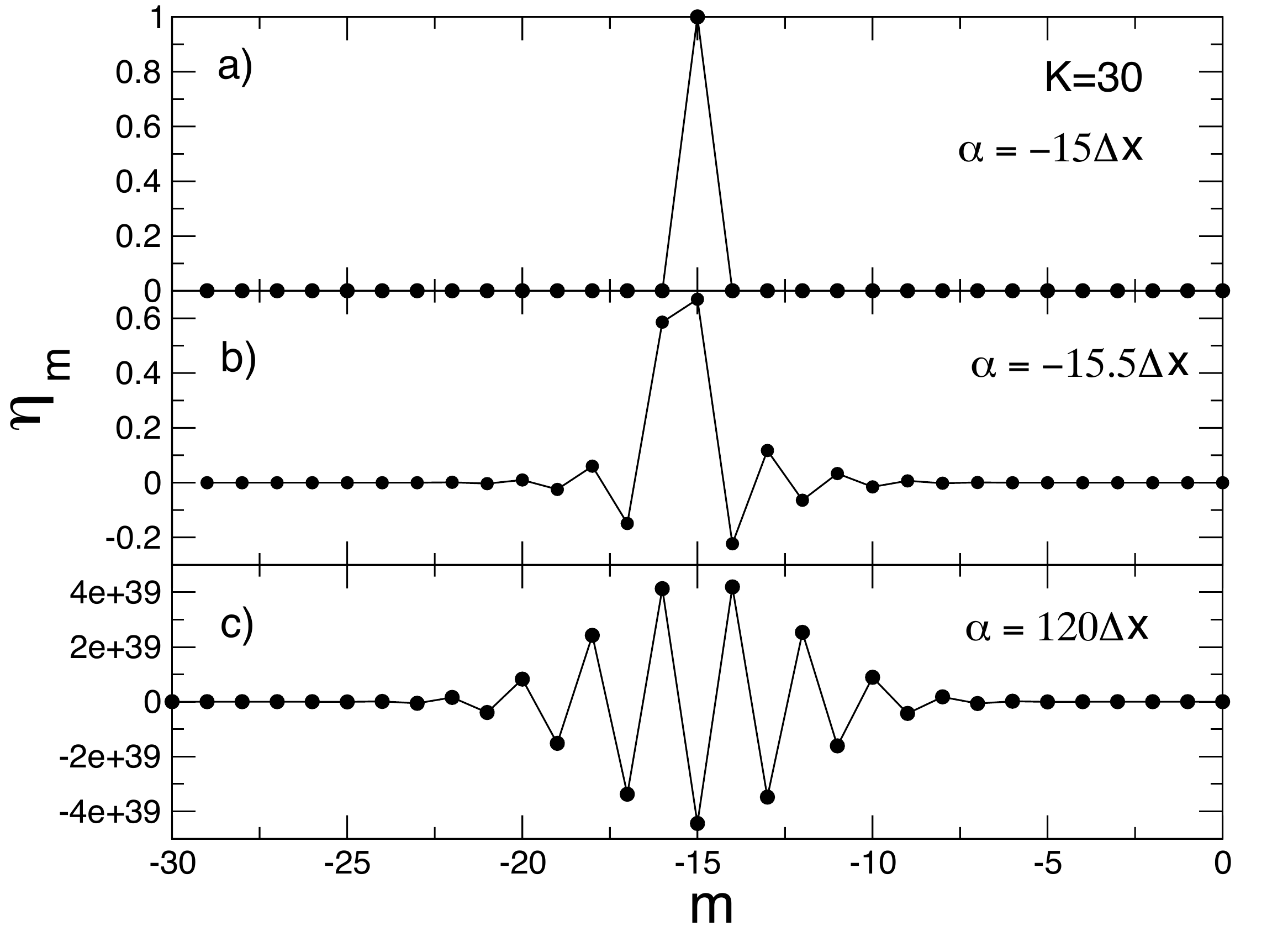}
\caption{The DAD (\ref{10})  for $K=30$ and $\a/\Delta x=-15$, $-15.5$ and $120$}.
\label{FIG2}
\end{figure}

{\it Coordinate space: quasi-Dirac distributions.} 
Equation (\ref{4}) requires  the delay amplitude distribution (DAD)   $\eta (x)$ in Eq.(\ref{6}) to be Dirac's $\delta(X-\a)$, something that cannot be achieved with the proposed set up. 
We can, however, use the freedom in choosing the
spin's initial and final states $|a\ra$ and $|b\ra$ to insure that the normalisation and the first $K$ moments of $\eta(x)$
equal those of $\delta (X-\a)$, 
\begin{eqnarray}\label{8}
\bar{x^n} \equiv \int_{-\infty}^\infty x^n \eta(x) dx=\a^n, \quad n=0,1,...K.
\end{eqnarray} 
This is equivalent to $K+1$ linear equations for the unknowns
$\eta_m$, 
\begin{eqnarray}\label{9}
\sum_{m=0}^{-K}A_{n,m}\eta_m=\a^n, \quad n=0,1,...K
\end{eqnarray} 
with a van der Monde matrix \cite{MONDE}
$A_{n,m} \equiv (m\Delta x)^n$.
Note that for $\eta_m \ge 0$ Eq. (\ref{8}) has a unique solution $\eta(x)=\delta(x-\alpha)$.
With no such restriction,
 other non-trivial solutions of Eqs.(\ref{9}) are possible.
Solving Eqs. (\ref{9}) in a usual manner \cite{FOOT1} yields (${\prod'} _{j=0}^K$ indicates   the  product
over all $j\ne m$)
\begin{equation}\label{10}
\eta_m(\a/\Delta x)=
(-1)^{m}\frac{{\prod'} _{j=0}^K(j+\a/\Delta x)}{m!(K-m)!}.
\end{equation}
With the quantities $\eta_m(\a)$ defined in Eq. (\ref{10}) we have a somewhat unusual mathematical object, an alternating distribution
 $\eta(x)$ with  a finite support inside $[-K\Delta x,0]$, which
acts as the Dirac's $\delta(x-\a)$ with support at an arbitrary $\a$, positive or negative, in the space of all polynomials of order no higher than
$K$. We may also expect it to have the same effect on any function $G(x)$ in a domain where its Taylor series can 
be truncated after $K$ terms,
\begin{eqnarray}\label{10a}
\int G(x) \eta(x) dx \approx 
\sum_{n=0}^K G^{(n)}/n! \int x^n \eta(x)dx = \\
\nonumber
\sum_{n=0}^K G^{(n)}\a^n/n!
\approx G(\a),
\end{eqnarray}
although to check the validity of Eq. (\ref{10a})  we need, in principle, also analyse the behaviour of $\bar{x^n}$ with $n>K$ (Fig.\ref{FIG1}).
We will refer to an $\eta(x)$ with the above properties as a {\it quasi-Dirac distribution} (of order $K$).
One example of such a distribution (of order 2) is the restricted path integral for the quantum traversal
time \cite{UNCERT}, which prompted Baz' \cite{BAZ} who compared its first and second moments, to conclude (erroneously \cite{UNCERT} )  that the duration of an elastic collision time has a unique 
well defined value.
We note further that a delay belonging to the discrete ``spectrum'' of available 
shifts, $-M\Delta x$, $-K \le M \le 0$,
is produced 
by choosing either $a_m=\delta_{mM}$
or $b_m=\delta_{mM}$, and Eq. (\ref{10}) yields, as it should, $\eta_m = \delta_{mM}$ (see Fig.\ref{FIG2}a).
Choosing a delay between two available shifts, say, $(M+1/2)\Delta x$,
produces an alternating distribution $\eta_m$ plotted in Fig.\ref{FIG2}b.
Finally, an attempt to achieve a significant speed up (delay)  
outside the support of $\eta(x)$, $|\a/K\Delta x| > 1$, yields $\eta_m$ which have
very large absolute values, yet sum to unity as required by the first of  Eqs. (\ref{9}).
The values $\eta_m$ for $\a/K\Delta x =4$ are shown in Fig.\ref{FIG2}c,  where we had to use 
David Bailey's multiple precision program \cite{BAIL} to compute them to 
a sufficiently high accuracy.
%
%

\begin{figure}[h]
\includegraphics[width=8.5 cm, angle=0]{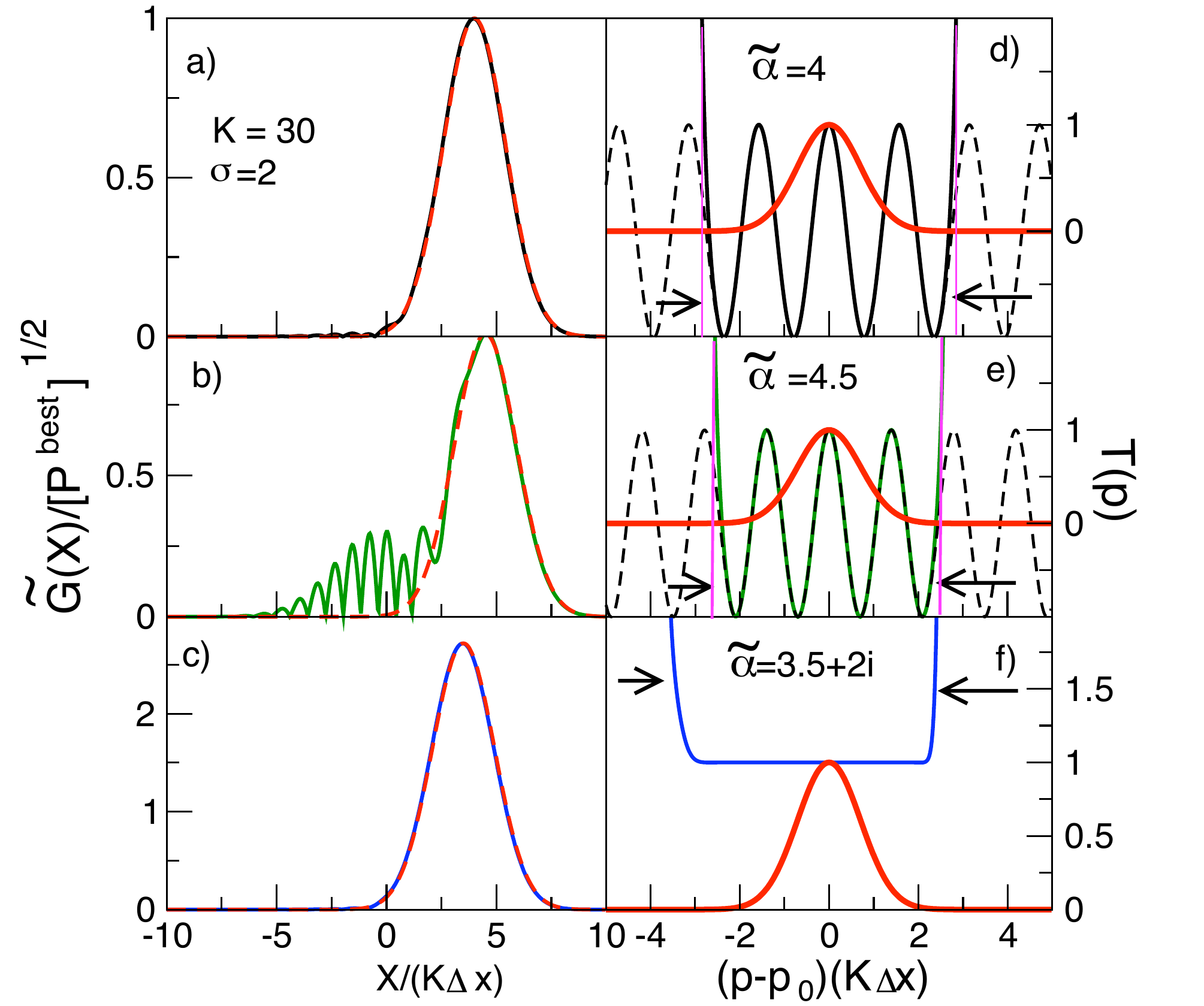}
\caption{a) The transmitted envelope (multiplied by the factor
$1/\sqrt{P^{best}}$ in Eq.(\ref{11})) for $K=30$, $\sigma/K\Delta x=2$ and $\tilde{\a}\equiv \a/(K\Delta x)=4.$
Also shown by the dashed line is $G(X-\a)$ in the r.h.s. of Eq.(\ref{4});
b) same as a) except for $\tilde{\a}=4.5$; c) same as a) except for 
$\tilde{\a}=3.5+2i$; 
d) $Re(T(p))$ vs. $p$ (solid) for the case shown in a). Also shown are $sin(-\a p)$ (dashed) and
$A(p)$ normalised to a unit height (thick solid). Vertical lines and arrows indicate the boundaries of the  supersocillatory window;
e) same as d) but for the case shown in b);
f) $|T(p)/exp(-i\a_1p+\a_2p)|$ (solid) and $A(p)$ (thick solid) for the case shown in c).}
\label{FIG3}
\end{figure}

{\it Probability of success and limits on the accuracy of transmission.}
Extremely large values of $\eta_m$  in  Fig.\ref{FIG2}c suggest that the proportion of the particles experiencing a significant advancement 
or delay will be very small as post-selection in the final state $|b\ra$ will be unlikely to succeed. 
Indeed, provided the envelope of the original pulse is reproduced without distortion (i.e., Eq. (\ref{4})
holds),  the norm of the transmitted state is given by $P(z)\approx 1/N(a)N(b)= [\sum_{m=-K}^0z_m \sum_{m=-K}^0|\eta_m|^2 /z_m]^{-1}$, 
$z_m\equiv |a_m|^2$ where we have used $|b_m|=|\eta_m|/|a_m|$, 
implied in Eq. (\ref{6a}). Maximazing $P(z)$ with respect to $z_m$, $m=-K,...0$ yields 
$|a_m|^2 =C |\eta_m|, \quad  |b_m|^2 = |\eta_m|/C, \quad $ where $C$ is a constant independent of $m$  so that the best probability of successful post-selection is given by
\begin{equation}\label{11}
  \quad   P^{best}(\a) = 1/(\sum_{m=-K}^0 |\eta_m|)^2.
  \end{equation}
As it has been shown above, for $|\a/K\Delta x| >1$, $\sum_{m=-K}^0 |\eta_m|$ is, unlike 
$\sum_{m=-K}^0 \eta_m =1$, a very large number, thus making  $P^{best}(\a)<<1$.
\newline
For given states $|a\ra$ and $|b\ra$, the success of advancing or delaying the free pulse depends 
on the shift $\a$ and the shape of the initial envelope $G(X)$, which we choose to be a Gaussian 
with a coordinate width $\sigma$,
\begin{equation}\label{12}
G(X) = (2/\pi \sigma^2)^{1/4} \exp(-X^2/\sigma^2).
\end{equation}
The shapes of the transmitted pulse are shown in Figs.\ref{FIG3} a,b and c. 
It is seen  that an increase of the shift $\a$ results in the rapid onset of oscillations
behind the peak of the transmitted pulse.

{\it Momentum space: superoscillations, real and complex-valued.}
To examine relation between the quasi-Dirac distributions of Sect.3 and
superoscillatory functions introduced in \cite {BERRY1} we expand the envelope $\tilde{G}(X)$  in plane waves,
%
\begin{eqnarray}\label{14}
\tilde{G}(X) = \int T(p)A(p)\exp(ipX)dp/ \sqrt {N(a)N(b)},
\end{eqnarray} 
where $G(x)=\int A(p)\exp(ipx)dp$ and the transmission amplitude  $T(p)$ is the Fourier transform of  the DAD $\eta(x)$,
\begin{equation}\label{15}
T(p) \equiv \int \eta(x) \exp(-ipx)dx = \sum_{m=0}^K \eta_m(\a) \exp(imp\Delta x).
\end{equation} 
%
 We note that  for
$\eta(x)=\delta (x-\a)$, Eq. (\ref{15}) would yield $T(p)=\exp(-ip\a)$.
With $\eta(x)$ being only a $K$-th order approximation to the Dirac $\delta(x-\a)$ we can expect such a behaviour in a limited range of $p$'s only.
For an advancement $\a > 0$, $T(p)$ which builds 
up from exponential with non-negative frequencies $m\Delta x$, $m=0,1...K$, will behave there as an 
exponential with a  negative frequency $-\a$, i.e., will be superoscillatory.
This superoscillatory window (band) is essentially the region around $p=0$ where both $T(p)$ and $\exp(-i\a p)$ are correctly
reproduced by the first $K$ terms of their Taylor series. Writing
$T(p) = \sum_{n=0}^K \partial^{(n)}_pT(0)p^n/n! $,
and recalling from Eqs.(\ref{8}) and (\ref{15}) 
\begin{eqnarray}\label{17}
 \partial^n_p T(0)= (-i)^k\bar{x^k}= (-i)^k \alpha^k, \quad 0\le k \le K
\end{eqnarray} 
we find that the $K$-th term is negligible provided $(|p||\alpha|)^K/K! << 1$, i.e, for,
\begin{eqnarray}\label{18}
|p| < K/(e|\a|),
\end{eqnarray} 
where we have used the Stirling formula for the factorial.
We can also choose the shift $\a$ to be complex valued,  $\a=\a_1+i\a_2$, in which case 
one with a complex frequency,
we have $T(p)\approx \exp(-ip\a_1+p\a_2)$, and find the free envelope in Eq.(\ref{4})
analytically continued into the complex coordinate plane.
In general, in order to avoid distortion we need  the momentum distribution of the initial wavepacket $A(p)$ to  fit into the  superoscillatory band of the transmission coefficient, beyond which  $T(p)$ exhibits rapid (polynomial) growth similar to that reported in Refs. \cite{BERRY1},\cite{AH1} (Fig.\ref{FIG3}). For a Gaussian envelope (\ref{12}) this requires (c.f. Eq.(\ref{18})) $2/\sigma << K/(e|\a|)$
which puts a restriction on the minimal  width of the wave packet which can be translated without distortion.

{\it Conclusions and discussion.}
In summary, we have proposed a system for which the transmitted state is a reduced copy
of the freely propagating one, translated along an arbitrary direction into the complex coordinate plane.
The shift is achieved by splitting the initial pulse into several copies delayed relative to free 
propagation. The system can be described in terms of a transmission amplitude  $T(p)$ or, alternatively,
by its Fourier transform, the delay amplitude distribution (DAD) $\eta(x)$. The DAD contains information about the spectrum of virtual delays experienced by the particle in the magnetic field and, therefore, about causality. For a real translation the DAD is a finite order approximation to a Dirac 
$\delta$-function and, for a complex valued shift, an analytic continuation of such an approximation. Accordingly, $T(p)$ exhibits
a superoscillatory band whose width limits the size of the shift and the width of the initial pulse.
The bandwidth depends on the number of first moments of the DAD 
(momentum derivatives of $T(p)$ at $p=0$) which satisfy $\bar{x^n} \approx \a^n$.
A similar analysis can be applied to any system described by a transmission amplitude which is analytic in the complex momentum plane, e.g., tunnelling across a potential barrier \cite{SMS}, 
where the shift is determined by the barrier width
and the DAD, like that in Eq. (\ref{6}) contains only delayed contributions \cite{SMS}. 

It is worth noting again that certain contradictions in the analysis of the  so-called tunnelling time problem \cite{REV} arise in an attempt to deduce the time $\tau$ a particle spends in the scatterer
from its delay $\Delta T = -\a/p_0$ in arriving at a remote detector.
This suggests that the time spent in the scatterer is $\tau=d/p_0-\a/p_0$ and the notion
of apparent ``superluminality'' arises if $\tau$ is made to be less than $d/c$.
 Our model readily demonstrates that there is no simple relation between $\Delta T$ and $\tau$. 
Indeed, between the field and the 
polariser the probability density corresponds, as it should, to $K$ real delays,
$P(x)=\sum_{m=-K}^0 |a_m|^2 |G(x-p_0 t-m\Delta x)|^2$. The advanced shape  (\ref{4}) is formed upon
passing a (possibly remote)  polariser, by an interference mechanism analogous to that responsible
for anomalous values obtained in weak measurements \cite{AHWEAK}.
There is also a similarity to the Wheeler's delayed choice experiment \cite{WHEEL}.
Just like Wheeler's photon may behave as a particle or as a wave, depending on which 
observation is made after the event,
the particle spends in the magnetic field a well defined duration $T=(d-m\Delta x)/p_0$
provided the polariser is set to select one of the states $|m\ra$.
If, on the other hand, the post-selection is performed in a linear combination of more than
one of such states, this duration cannot be defined and the experiment probes the wave 
nature of the transmitted particle.

{\it Acknowledgements:} DS and RSM are grateful to Max-Plank-Institute for Physics of Complex Systems
(Dresden) for hospitality and financial support; RMS acknowledges 
 Ministerio de Educaci\'on y Ciencia, 
Plan Nacional for support under grant No. FIS2007-64018.


\begin{thebibliography}{999}
\bibitem{REV} E. H. Hauge and J. A. Stoevneng, Rev. Mod. Phys. \textbf{61}, 917 (1989); 
C. A. A. de Carvalho, H. M. Nussenzweig, Rev. Mod. Phys. \textbf{364}, 83 (2002); 
V. S. Olkhovsky, E. Recami and J. Jakiel, Rev. Mod. Phys. \textbf{398}, 133 (2004).
\bibitem{Chen} Y.P.Wang and D.L. Zhang, Phys. Rev .A  \textbf{52}, 2597 (1995);
Y.Japha and G.Kurizki,  Phys. Rev .A  \textbf{53}, 586 (1996);
X.Chen and C.F.Li, Eur.Phys.Lett,  \textbf{82}, 30009 (2008).


\bibitem{BERRY1} M. V. Berry, J. Phys. A, {\bf 27}, L391 (1994).

\bibitem{KEMPF} A. Kempf, J. Math. Phys., {\bf 41}, 2360 (2000); 
A. Kempf and P.J.S Ferreira, J. Phys. A: Math. Gen., {\bf 37}, 12067 (2004);
M. S. Calder and A. Kempf, J. Math. Phys., {\bf 46}, 012101 (2005);
P. J. S .Ferreira, A. Kempf and M. J. C. S. Reis, J. Phys. A, {\bf 27}, L391 (2007).

\bibitem{BERRY2} M. V. Berry and S. Popescu, J. Phys. A, {\bf 39}, 6965 (2006);
F. M. Huang, Y. Chen, F. J. Garc\'\i a de Abajo and N. I. Zheludev, J. Opt. A: Pure Appl. Opt. {\bf 9}, S285 (2007);
M. R. Dennis, A. C. Hamilton and J. Courtial, Opt. Lett., {\bf 33}, 2976 (2008);
N. I. Zheludev, Nature (materials),{\bf 7}, 420 (2008);
M. V. Berry and M. R. Dennis, J. Phys. A: Math. Theor. {\bf 42}, 022003 (2009).

\bibitem{AH1} Y. Aharonov, N. Erez and B. Reznik, Phys. Rev. A, {\bf 65}, 052124 (2002).

\bibitem{SMS} D. Sokolovski, A. Z. Msezane and V. R. Shaginyan, Phys. Rev. A, {\bf 71}, 064103 (2005).

\bibitem{TOLL} J. Tollaksen, J. Phys.: Conf. Series, {\bf 70}, 012016 (2007).

\bibitem{MONDE} A. Ostaszewski, {\em Advanced Mathematical Methods } (Cambridge University Press, 1990).

\bibitem{FOOT1} 
Consider the determinant $\Delta_m (\alpha)$ of the matrix $A$ in which 
the $m$-th column is replaced by $(1,\alpha,\alpha^2,...\alpha^K)$. $\Delta_m (\alpha)$, which is a polynomial
of order $K$ in $\alpha$, must vanish for any $\alpha = -n\Delta x$, $n\ne m$, and, therefore, can be written 
as $\Delta_m (\alpha) =\prod_{n\ne m}(\alpha + n\Delta x)\times const$.
Also, we have $det A = \Delta_j(-j\Delta x)$, for any $0 \le j \le K$, and the ratio
$\Delta_m (\alpha)/det A$ yields Eq.(\ref{10}).

\bibitem{UNCERT} D. Sokolovski, Phys. Rev. A, {\bf 76}, 042125 (2007).

\bibitem{BAZ} A. I. Baz': Yad. Fiz. \textbf{4}, 182 (1967) [Sov.J.Nucl.Phys, {\bf 5}, 635 (1967)].

\bibitem{BAIL} D. H. Bailey, ACM Transactions on Mathematical Sotfware, {\bf 19}, 288 (1993).

\bibitem{AHWEAK} Y. Aharonov, D. Albert and L. Vaidman, Phys. Rev. A, {\bf 60}, 1351 (1988).

\bibitem{WHEEL} J. A. Wheeler, in {\it Mathematical Foundations of Quantum Theory}, edited by A. R. Marlow, Academic Press, New York, 1978, p. 13.


\end{thebibliography}
\end{document}